\documentclass[twoside,fleqn]{ActaStyle}
\usepackage{times,cite}

\usepackage[dvips]{graphicx}    % if you need to include figures in PostScript
\usepackage[tbtags]{amsmath}    % if you need more math

\def\authorheading{E. Oset, J. R. Pel\'aez, L. Roca}

\def\shorttitle{$\eta \to \pi^0 \gamma \gamma$ decay}

\begin{document}
%% ---------------------------------------------------

%%%  page range, first and last page
\pagerange{1}{6}

%%% paper title
\title{Discussion of the $\eta \to \pi^0 \gamma \gamma$ decay within a 
chiral unitary approach
}

%%% author(s) and address(es)
\author{
E. Oset\email{oset@ific.uv.es}$^*$, J. R. Pel\'aez \,$^\dagger$, L. Roca\,$^*$,
}
{$^*$ Departamento de F\'{\i}sica Te\'orica and IFIC
 Centro Mixto Universidad de Valencia-CSIC\\
 Institutos de Investigaci\'on de Paterna, Apdo. correos 22085,
 46071, Valencia, Spain\\
 }
$^\dagger$ Departamento de
F\'{\i}sica Te\'orica II,  Universidad Complutense. 28040 Madrid,
Spain\\

%%% Date of submition
\day{April 5, 2002}

%%% abstract of the paper
\abstract{We improve the calculations of the  $\eta \to
\pi^0 \gamma \gamma$ decay within the context of meson chiral
lagrangians. We use a chiral unitary approach for the meson-meson 
interaction, thus generating the $a_0(980)$ resonance and fixing
the longstanding sign ambiguity on its contribution. This also
allows us to calculate the loops with one vector meson exchange,
thus removing a former source of uncertainty.  In addition we
ensure the consistency of the approach with other processes.
First, by using vector meson dominance couplings normalized to
agree with radiative vector meson decays. And, second, by checking
the consistency of the calculations with the related $\gamma
\gamma \to \pi^0 \eta$ reaction. We find an $\eta \to \pi^0 \gamma
\gamma$ decay width of $0.47\pm 0.10$ eV,  in clear disagreement
with published data but in remarkable agreement with the most
recent measurement.
}

%%% PASC numbers of your article

% %%%%%%%%%%%%%%%%%%%%%%%%%%%%%%%%%%%%%%%%%%%%%%%%%%%%%%%%%%%%%%%%%
\section{Introduction}
% %%%%%%%%%%%%%%%%%%%%%%%%%%%%%%%%%%%%%%%%%%%%%%%%%%%%%%%%%%%%%%%%%
The $\eta \to \pi^0 \gamma \gamma$ decay has attracted much
theoretical  attention, since Chiral Perturbation Theory (ChPT)
calculations have sizable uncertainties and  produce
systematically rates  about a factor of two smaller than
experiment. Within ChPT, the problem stems from the fact that the
tree level amplitudes, both at $O(p^2)$ and $O(p^4)$, vanish. The
first non-vanishing contribution comes at $O(p^4)$, but either
from loops involving kaons,  largely suppressed due to the kaon
masses, or from pion loops, again suppressed since they violate G
parity and are thus proportional to $m_u -m_d$.  The first sizable
contribution comes at  $O(p^6)$ but the coefficients involved are
not precisely determined. The use of tree level VMD  to obtain
the $O(p^6)$ chiral coefficients by expanding the vector meson
propagators, leads  to results about a
factor  of two smaller than the "all order" VMD term, which means
keeping the full vector meson propagator. All this said it has become clear 
that the
strict chiral counting has to be abandoned since the $O(p^6)$  and
higher orders involved in the full (``all order'') VMD results are
larger than those of $O(p^4)$. 

Once the ``all order'' VMD results is accepted as the dominant
mechanism, one cannot forget the tree level exchange of other
resonances around the 1 GeV region. The  $a_0(980)$ exchange,
which was taken into account approximately in 
\cite{Ametller:1991dp},  was one of the main sources of
uncertainty, since even the sign of its contribution was unknown.

After the tree level light resonance exchange has been 
taken into account, we should consider loop diagrams,
since meson-meson interaction or rescattering can be rather strong. First
of all we find the already commented $O(p^4)$ kaon loops from ChPT,
but also the meson loops from the terms involving
the exchange
of one resonance. The uncertainty from the latter was roughly 
expected \cite{Ametller:1991dp} to be about 30\% of the full width.

 Another relevant question is that no attempts have been done to
check the consistency of $\eta \to \pi^0 \gamma \gamma$ results
with the related channel $\gamma \gamma\to \pi^0 \eta$.  The
reason is not surprising since there are no hopes within ChPT to
reach the $a_0(980)$ region where there are measurements of the
$\gamma \gamma\to \pi^0 \eta$ cross section. On the other hand,
the explicit SU(3) breaking already present in the  radiative
vector meson decays has not been taken into account when
calculating the VMD tree level contributions, and this effect changes 
the results obtained from VMD estimations by about a factor of two.

The former discussion has set the stage of the problem and the 
remaining uncertainties that allow for further improvement.
In recent years, with the
advent of unitarization methods, 
it has been possible to extend the results of 
ChPT to higher energies where  the perturbative expansion
breaks  down and to generate resonances up to 1.2 GeV. 
In particular these
ideas were used to describe the $\gamma \gamma \to meson-meson$ reaction, with
good results in all the channels up to energies of around 1.2 GeV 
\cite{Oller:1997yg}.
With these techniques,
and always within the context of meson chiral lagrangians,
 we will address three of the problems stated above:
First,  the $a_0(980)$ contribution, second,  the evaluation of meson loops
from VMD diagrams and, third, the consistency with the  
crossed channel $\gamma \gamma\to \pi^0 \eta$.
In particular, we will make use of the results
in  \cite{Oller:1997yg}, where the $\gamma \gamma\to \pi^0 \eta$
cross section around the $a_0(980)$ resonance  
was well reproduced  using the 
same input as in 
meson meson scattering, without introducing any extra parameters.

With these improvements we are then left with a model
that includes the ``all order'' VMD and resummed chiral loops.

\section{Mechanisms}

\begin{figure}
\begin{center}
  \includegraphics[height=.08\textheight]{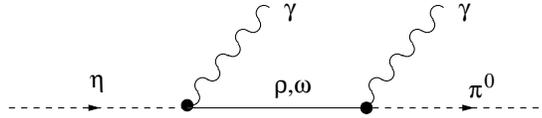}
  \caption{Diagrams for the VMD mechanism.}
 \end{center}
\end{figure}

Following \cite{Ametller:1991dp} we consider the sequential VMD
mechanism of Fig.~1
which can  be easily derived from the VMD Lagrangians involving
VVP and $V\gamma$ couplings
\begin{equation}
{\cal L}_{VVP} = \frac{G}{\sqrt{2}}\epsilon^{\mu \nu \alpha \beta}\langle
\partial_{\mu} V_{\nu} \partial_{\alpha} V_{\beta} P \rangle, \qquad
{\cal L}_{V \gamma} =-4f^{2}egA_{\mu}\langle QV^{\mu}\rangle,
\label{lagr}
\end{equation}
where $V_{\mu}$ and $P$ are standard $SU(3)$ matrices
constructed with the nonet of vector mesons containing the
$\rho$, and the nonet of pseudoscalar mesons containing the
$\pi$, respectively. 
 We also assume an ordinary mixing for the $\phi$,
the $\omega$, the $\eta$ and $\eta'$.

From Eq.~(\ref{lagr}) one can obtain the radiative widths for
$V\to P\gamma$ obtaining a fair agreement with the experimental
data in the PDG but the results can be improved by incorporating
$SU(3)$  breaking mechanisms. For that purpose, {\it we  normalize 
the  couplings so that the  branching ratios  agree with
experiment}. These will be called results with ``normalized
couplings''. In this way we are taking into account 
phenomenologically the corrections to the $VP\gamma$ vertex from
an underlying field theory.

The integrated width obtained using this sequential VMD
contribution is 
$\Gamma=0.57\,$eV (universal couplings); 
$\Gamma=0.30 \pm 0.06\,$eV (normalized couplings),
where the error has been calculated from a Monte Carlo Gaussian sampling
of the normalization parameters within the errors of the experimental 
branching ratios.

Our  VMD normalized
 result is within three standard deviations from the value presently given
in the PDG:  $\Gamma=0.84 \pm 0.18\,$eV,
but within one sigma of the more recent one 
presented in \cite{nefkens}, $\Gamma=0.42 \pm 0.14\,$eV.
There are, however, other contributions that we consider next.

The contribution of pion loops to $\eta \to \pi^0 \gamma \gamma$,
evaluated  in \cite{Ametller:1991dp},
proceeds, to begin with, through the  G-parity violating $\eta\to
\pi^0\pi^+\pi^-$ process but it is
proportional to $m_u-m_d$, 
and we shall include it in the uncertainties of
together with other isospin violating contributions.

The main meson loop contribution comes from  the charged kaon 
loops at  $O(p^4)$ and proceeds via $\eta \to \pi^0 K^+
K^-\rightarrow \pi^0\gamma\gamma$. Note that these loops are also
suppressed due to the large kaon masses. That is why the 
$\eta\rightarrow\pi^0 a_0(980)\rightarrow\pi^0\gamma\gamma$
mechanism  was included explicitly, with uncertainties in the size
and sign of the $a_0(980)$ couplings. As commented in the
introduction, the chiral unitary approach solves this problem by
generating dynamically the $a_0(980)$ in the $K^+ K^-\rightarrow
\pi^0\eta$ amplitude.

We can illustrate this approach 
by revisiting the work done in \cite{Oller:1997yg} on the
related process $\gamma\gamma\to
\pi^0\eta$ where  the chiral unitary approach was successfully
applied around the $a_0(980)$ region. 
Since for the $\eta$ decay the low energy region of
$\gamma\gamma\to
\pi^0\eta$ is also of interest, we will include
next the VMD mechanisms also in this reaction. 
Once we check that we describe correctly
$\gamma\gamma\to
\pi^0\eta$, the results can be easily translated
to the eta decay. We will finally add other 
anomalous meson loops that are numerically relevant for eta decay but not 
for $\gamma\gamma\to
\pi^0\eta$.

In \cite{Oller:1997yg} it was shown that,
within the unitary chiral approach, the $\gamma\gamma\to\pi^0\eta$ amplitude 
around the $a_0(980)$ region, diagrammatically represented at one
loop in Fig.~2,
factorizes as
\begin{equation}
-it=(\tilde{t}_{\chi K}+ \tilde{t}_{AK^+K^-}   )t_{K^+K^-,\pi^0\eta}
\label{eq:tNPA629}
\end{equation}
with $t_{K^+K^-,\pi^0\eta}$ the full $K^+K^-\to\pi^0\eta$ transition
amplitude.

\begin{figure}
\begin{center}
  \includegraphics[height=.30\textheight]{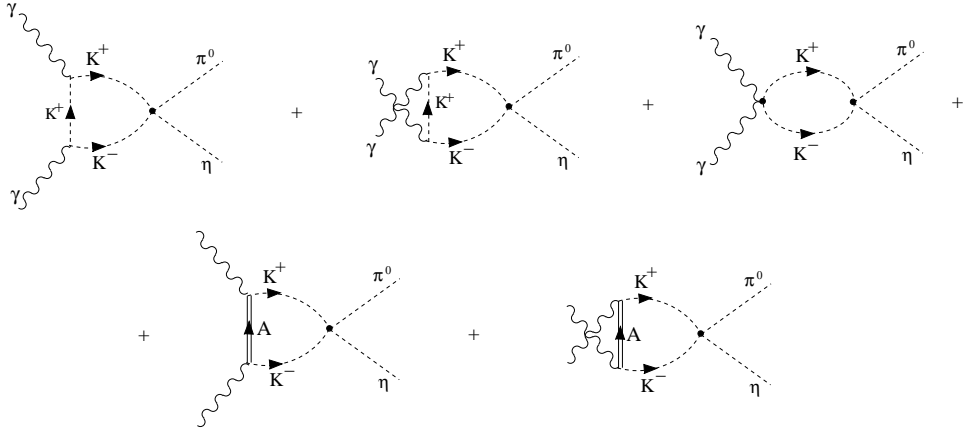}
  \caption{Diagrams for the chiral loop contribution.}
  \end{center}
\end{figure}

The first three diagrams correspond to $\tilde{t}_{\chi K}\,
t_{K^+K^-,\pi^0\eta}$ of Eq.~(\ref{eq:tNPA629}). 
The meson meson scattering amplitude was evaluated in
\cite{Oller:1997ti} by summing
the Bethe Salpeter (BS) equation with a kernel formed from the
lowest order meson chiral Lagrangian amplitude and
regularizing the loop function with a three momentum cut off.
Other approaches like the inverse amplitude
method or the $N/D$ method all
give the same results in the meson scalar sector. 
The BS equation with coupled channels can be 
 solved algebraically, leading to the following solution in matrix form 
\begin{equation}
t(s)=[1-t_2(s)G(s)]^{-1}t_2(s),
\end{equation}
with $s$ the invariant mass of the two mesons, $t_2$ the lowest
order chiral amplitude and $G(s)$ a diagonal matrix,
$\mbox{diag}(G_{\overline{K}K},G_{\eta\pi})$, accounting for the loop
functions of two mesons. These $G$ functions were regularized in
\cite{Oller:1997ti}
by 
means of a cut off.

In Eq.~(\ref{eq:tNPA629}) there is another term, $\tilde{t}_{AK^+K^-}
t_{K^+K^-,\pi^0\eta}$, which corresponds to the last two
diagrams of Fig.~3 where the axial vector meson $K_1(1270)$
 is exchanged.

In addition to the axial vector meson
exchange in loops considered,
we have to include the loops with vector meson exchange for completeness.
In fact, some of the uncertainties estimated in 
\cite{Ametller:1991dp} were linked to these loops. 
For consistency, once again we have to sum
the series obtained by iterating the loops in the four meson
vertex.

Of course, when introducing loops with vector meson exchange
we have to consider loops involving 
a $K^{*+}$
or a $K^{*0}$ exchanged between the photons,
which were not present at tree level.

In the $\eta\to\pi^0\gamma\gamma$ case, the meson loop diagrams
correspond to those of $\pi^0\eta\to\gamma\gamma$ but
considering the $\pi^0$ as an outgoing particle.

Since we are considering
all the VMD diagrams and the chiral loops, 
we still have to take into account
another kind of loop diagrams
\cite{Ametller:1991dp} which involve two anomalous 
$\gamma\rightarrow 3 M$ vertices.
Despite being  $O(p^8)$ 
it has been  found \cite{Ametller:1991dp} that they can have a
 non negligible effect on the $\eta$ decay.

\begin{figure}
\begin{center}
  \includegraphics[height=.29\textheight]{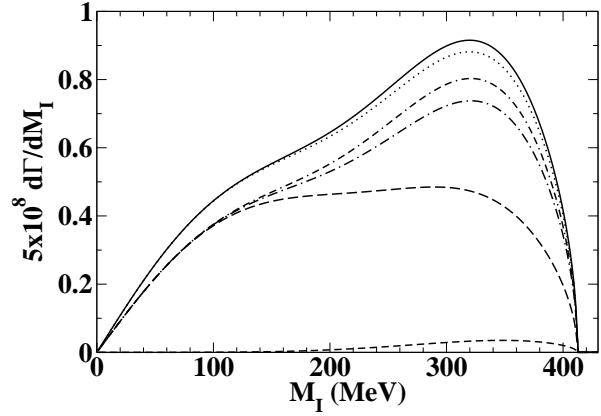}
  \caption{Contributions to the two photon invariant mass distribution.
 From bottom to top,
short dashed line: chiral loops from Eq.(\ref{eq:tNPA629}); 
long dashed line: only tree level VMD;
dashed-dotted line:
coherent sum of the previous mechanisms; double
dashed-dotted line: idem  but adding the resummed VMD loops;
continuous line: idem but adding the anomalous terms,
which is the full model presented in this work 
(we are also showing as a dotted line the
full model but substituting the full 
$t_{K^+K^-,\eta\pi^0}$
amplitude by its lowest order).}
\label{fig:ggexp}
\end{center}
\end{figure}

\section{Results}

Using the model described in the previous section, we plot in
Fig.~4 the different contributions to $d\Gamma/dM_I$. We can see
that the largest contribution is that of the  tree level VMD (long
dashed line).  Let us recall that this is a new result as long as 
we are using the VMD couplings normalized to  agree with the
vector radiative decays. The resummation of the loops in Fig.~2
using Eq.(\ref{eq:tNPA629}), (short dashed line) gives a small
contribution ($0.011\,$eV in the total width),  but when added
coherently to the tree level VMD,  leads to an increase of
$30\,$\% in the $\eta$ decay rate (dashed-dotted line). More
interestingly, the shape of the $\gamma\gamma$  invariant mass
distribution is appreciably changed with respect to the tree level
VMD, developing a peak at high invariant masses. The  resummed VMD
loops leads, through interference, to a moderate increase of the
$\eta$ decay rate (double dashed dotted line), smaller  than that
of the chiral loops considered before. The last ingredient is the
contribution of the anomalous mechanisms (continuous line), 
leading again to a moderate increase of the $\eta$ decay rate,
also smaller than  the chiral loops from Eq.(\ref{eq:tNPA629}).
These anomalous mechanisms have a very similar shape to the tree
level VMD and interfere with it in the whole range of invariant
masses. Altogether the final result is

\begin{equation}
\Gamma(\eta\to\pi^0\gamma\gamma)=0.47\pm 0.10\,eV
\label{resultfinal}
\end{equation}
where the theoretical error  have been obtained considering the
uncertainties from the vector meson radiative decays, the
contribution of the $1^{+-}$ axial-vector mesons and the isospin
violating terms.

Note that although we have considered a new error source 
from the uncertainties in the vector radiative decays,
which turns out to be the largest one,
we still have reduced the uncertainty from previous calculations.

The result of Eq.~(\ref{resultfinal}) is in remarkable agreement 
with the latest experimental number $\Gamma=0.42 \pm 0.09\ \pm 0.08$ $eV$ 
\cite{nefkens},
based on 1600 events and also with those of \cite{knecht} of 
$\Gamma=0.35 \pm 0.20$ based on a smaller statistics of about 120 events.
and lie within two sigmas from the 
 earlier one in the PDG $\Gamma=0.84 \pm 0.18$  $eV$. Confirmation
of those preliminary results  would therefore be important to test
the consistency of this new approach. Furthermore,
precise measurements of  
the $\gamma\gamma$ invariant mass distributions would 
be of much help given the differences found
 with and without loop contributions.

Finally, we would like to make some precisions concerning the comparison of
recent experimental data with former calculations of chiral perturbation
theory plus VMD estimates. 
The chiral models prior to our work  have to
be looked in perspective.  They rely upon data on radiative decay of
vector mesons which has changed considerably in recent years.  We use updated
data and find that if the earlier calculations would have used the present data
they would get eta decay widths about one half what they got.  
Thus, comparing our result and the old ones is somewhat misleading. And it is
also misleading to compare the experiments with these calculations without this
warning. We find also inappropriate to compare experimental results with what
chiral perturbation theory would give at order $O(p^6)$ since, as we have
mentioned here, these results are obtained from a VDM model projecting over 
$O(p^6)$ the results of the full model, which provides a width about a factor
two larger than its  $O(p^6)$ projection. In other words, a strict chiral perturbative
calculation in terms of powers of momentum is not practical for this problem,
and the full model predictions, which accounts also for higher powers of
momentum, have to be taken for reference.  

  In addition, previous works in the literature have
large uncertainties from ignorance of the $a_0(980)$ contribution.  
We have improved all this and
other small things and evaluate theoretical errors from different sources. 
In summary the most relevant things of the present work  are:

1) Use of updated data for radiative decay of vector mesons.  If previous chiral
calculations are updated making use of these new data
the result comes about a factor of two smaller.

2) Unitarized chiral perturbation theory allows to determine precisely the
contribution of the $a_0(980)$ resonance, which was a major source of uncertainty
in the past.

3) A careful determination of theoretical errors has been done, which was also
lacking before.

Our work has made the most accurate calculation, so far, withing the framework
of chiral dynamics and can be taken as reference of what chiral theory predicts
for this reaction.  Previous chiral works have to be seen in their value as
making the first predictions for this ratio, but should not be used as a
reference for a quantitative prediction since they contain intrinsic
uncertainties of more than a factor two.\\

{\bf Acknowledgments:}\ \ 
This work is partly supported by DGICYT contracts BFM2002-01868, 
and the E.U. EURIDICE network contract HPRN-CT-2002-00311.
This research is part of the EU Integrated Infrastructure Initiative
Hadron Physics Project under contract number RII3-CT-2004-506078.

\end{document}